\documentclass[twocolumn,prl,showpacs,floatfix,amsmath,amsfonts]{revtex4}
\usepackage[latin9]{inputenc}
\usepackage{color}
\usepackage{amsmath}
\usepackage{graphicx}
\usepackage{esint}

\makeatletter
\@ifundefined{textcolor}{}
{%
 \definecolor{BLACK}{gray}{0}
 \definecolor{WHITE}{gray}{1}
 \definecolor{RED}{rgb}{1,0,0}
 \definecolor{GREEN}{rgb}{0,1,0}
 \definecolor{BLUE}{rgb}{0,0,1}
 \definecolor{CYAN}{cmyk}{1,0,0,0}
 \definecolor{MAGENTA}{cmyk}{0,1,0,0}
 \definecolor{YELLOW}{cmyk}{0,0,1,0}
}

\usepackage{color}
\usepackage{graphics}
\usepackage{epsfig}
\newcommand{\be}{\begin{equation}}
\newcommand{\ee}{\end{equation}}
\newcommand{\bea}{\begin{eqnarray}}
\newcommand{\eea}{\end{eqnarray}}
\newcommand{\bes}{\begin{subequations}}
\newcommand{\ees}{\end{subequations}}
\newcommand{\PT}{\mathcal{PT}}

\newcommand{\p}{{\cal P}}
\newcommand{\T}{{\cal T}}

\newcommand{\bpsi}{ \mbox{\boldmath$\psi$\unboldmath}  }

\newcommand{\bu}{\mathbf{u}}

\newcommand{\bPsi}{\mathbf{\Psi}}

\newcommand{\bc}{{\bf c}}

\makeatother

\begin{document}

\title{Bloch Oscillations in Optical and Zeeman Lattices in the Presence
	of Spin-Orbit Coupling}

\author{Yaroslav V. Kartashov$^{1,2}$, Vladimir V. Konotop$^{3}$, Dmitry
A. Zezyulin$^{3}$, and Lluis Torner$^{1,4}$ }

\affiliation{$^{1}$ ICFO-Institut de Ciencies Fotoniques, The Barcelona Institute
of Science and Technology, 08860 Castelldefels (Barcelona), Spain
\\
 $^{2}$Institute of Spectroscopy, Russian Academy of Sciences, Troitsk,
Moscow Region, 142190, Russia \\
 $^{3}$ Centro de F\'{i}sica Teórica e Computacional, Faculdade de
Ciências and Departamento de F\'{i}sica, Faculdade de Ciências, Universidade
de Lisboa, Campo Grande, Ed. C8, Lisboa 1749-016, Portugal}

\affiliation{$^{4}$Universitat Politecnica de Catalunya, 08034 Barcelona, Spain }

\date{\today}
\begin{abstract}
We address Bloch oscillations of a spin-orbit coupled atom in
periodic potentials of two types: Optical and Zeeman lattices. We
show that in optical lattices the spin-orbit coupling allows controlling
the direction of atomic motion and may lead to complete suppression
of the oscillations at specific values of the coupling strength. In
Zeeman lattices the energy bands are found to cross each other at
the boundaries of the Brillouin zone, resulting in period-doubling
of the oscillations. In all cases, the oscillations are accompanied by rotation
of the pseudo-spin, with a dynamics that is determined by the strength of
the spin-orbit coupling. The predicted effects are discussed also in
terms of a Wannier-Stark ladder, which in optical lattices consist
of two mutually-shifted equidistant sub-ladders.
\end{abstract}


\maketitle
%%%%%%%%%%%%%%%%%%%%%%%%%%%%%%%%%%%%%%%%%%%%%%%%%%%%%%%%%%%%%%%%%%%

Bloch oscillations (BOs) of electrons in a crystal under the action
of a constant electric field~\cite{Zener} is one of the fundamental
predictions of quantum mechanics. They were observed
in semiconductor superlattices~\cite{experim_solid,Leo},
a few years after the observation~\cite{WSL} of a Wannier-Stark
ladder (WSL). BOs have been observed
in a variety of systems where waves can propagate in
periodic environment~\cite{reviews}, such as cold atoms~\cite{Peik}, Bose-Einstein condensates (BECs)~\cite{BEC}
held in optical lattices, and diverse optical settings, including
waveguide arrays~\cite{Peshel}, optically-induced lattices \cite{key-3},
coupled microcavities~\cite{Sapienza}, or plasmonic systems \cite{key-2}.

To date, most studies about BOs in atomic systems have addressed one-component settings. Two-component systems have received less attention.
A particularly interesting situation occurs in the presence of linear
coupling between components that in steady state locks their energies. Spin-orbit (SO) coupling can be implemented in
a multi-level atom~\cite{multi-level} for which artificial electric~\cite{electric}
and magnetic~\cite{magnetic,magnetic_1} fields can be created by
managing interactions between different hyperfine states~\cite{gauge}.
A system involving coupled hyperfine atomic states in a periodic
potential and subjected to linear forces induced by a Zeeman field undergo BOs~\cite{Witthaut} in a static field
or quantum walks in a time-periodic field. The dynamics of such a system
is strongly influenced by the presence of two oppositely
tilted WSL in the eigenmode spectrum~\cite{Ke}. Interesting dynamical regimes
occur in SO-coupled BECs~\cite{Spielman}. The interplay of
BOs due to a linear force and the spin Hall effect induced by the SO coupling
can lead to  complex  evolution of BECs
in a two-dimensional lattice~\cite{Larson}. {\color{black} Both, BOs and WSL of a four-component model of a helicoidal molecule with SO coupling have been addressed~\cite{Caetano}}.

In this Letter we report unconventional features
that SO coupling brings into the dynamics of BOs in a
system held either in an optical lattice (OL) or in a Zeeman lattice (ZL) and subjected to
a linear force. We highlight suppression of
BOs due to band-flattening in optical lattices, transition
from periodic to non-periodic BOs, period-doubling
of BOs in Zeeman lattices due to band-crossing at the
edges of the Brillouin zone (BZ), and control of the direction of motion of broad wavepackets.

\begin{figure}
\begin{centering}
\includegraphics[width=1\columnwidth]{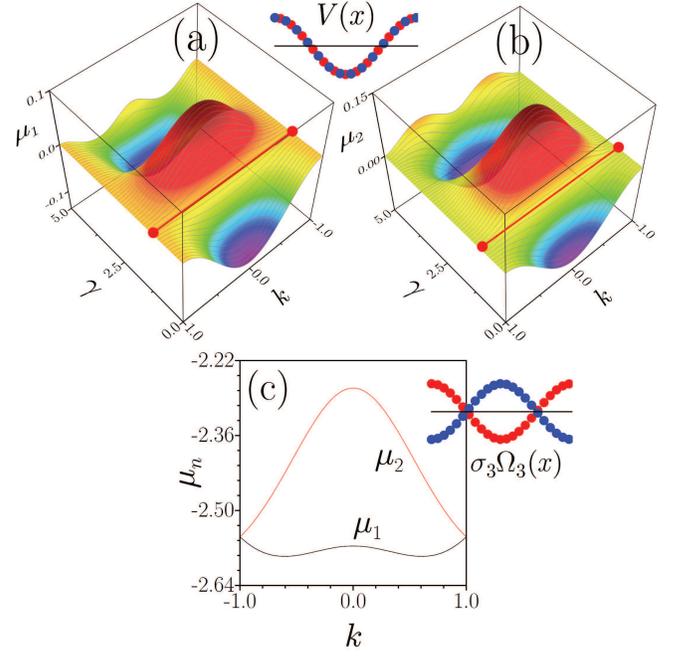}
\par\end{centering}

\caption{(Color online) First (a) and second (b) bands
in the OL spectrum as functions of $\gamma$. A vertical shift of
the bands with increasing $\gamma$ was eliminated by subtracting
$\mu_{n}(k=1)$ from $\mu_{n}(k)$. The red lines indicate some of the
points where band flattening occurs. (c) First two bands in the spectrum
of ZL at $\gamma=2$. The top and bottom insets show OL and ZL and the blue and red dots correspond to the potentials affecting different components.}
\label{fig:one}
\end{figure}

We consider a two-level atom described by the spinor $\boldsymbol{\Psi}=(\Psi_{1},\Psi_{2})^{T}$,
which solves the Schr\"odinger equation   $i\mathbf{\boldsymbol{\Psi}_{\mathit{t}}}=(H_{0}+H_{L}+\beta x)\mathbf{\boldsymbol{\Psi}}$, {\color{black} where time and position are measured respectively in units of $2m d^2/(\hbar\pi^2)$ and $d/\pi$, with $m$ being the mass of the atom and $d$ the lattice period}.   $H_{0}=p^{2}+\gamma\sigma_{3}p+\Omega_{1}\sigma_{1}$ is the
Hamiltonian without lattice and external field, $p=-i\partial/\partial x$
is the momentum, $\gamma$ is the SO coupling strength, $\Omega_{1}$ is the Rabi frequency, $\beta$ is the linear
force, and $\sigma_{1,2,3}$ are Pauli matrices. The potential
$H_{L}(x)$ is set by an OL $H_{L}(x)=V(x)$ that is equal for
both spinor components, or by a ZL  $H_{L}(x)=\Omega_{3}(x)\sigma_{3}$ having opposite signs for the components. Both lattices are $\pi$-periodic, i.e., $H_{L}(x)=H_{L}(x+\pi)$;
even functions with respect to the origin $[H_{L}(x)=H_{L}(-x)]$; and odd functions with
respect to the quarter-period $[H_{L}(\pi/4+x)=-H_{L}(\pi/4-x)]$. In simulations we model OLs and ZLs as $V(x)=-4\cos(2x)$ and $\Omega_3(x)=4\cos(2x)$, respectively [see Fig.~\ref{fig:one}].
Both such potentials are experimentally feasible as described in Refs.~\cite{BEC} and \cite{Zeeman} for OLs  and ZLs, respectively.

When $\beta=0$, the eigenmodes of the system are Bloch waves $\boldsymbol{\Psi}=\boldsymbol{\psi}_{k}e^{-i\mu(k)t}$,
i.e., $(H_{0}+H_{L})\boldsymbol{\psi}_{k}=\mu(k)\boldsymbol{\psi}_{k}$,
where $\boldsymbol{\psi}_{k}=\mathbf{u}_{k}(x)e^{ikx}$, $\mathbf{u}_{k}(x)=\mathbf{u}_{k}(x+\pi)$,
and $k$ is the Bloch momentum in the reduced BZ: $k\in[-1,+1]$. {\color{black} We start with a semi-classical description valid for} broad wavepackets {\color{black} (namely, $\delta k d\ll 1$, where $\delta k$ is the spectral width of the wavepacket)}. Similarly to  the one-component case~\cite{Zener,Houston}, one can show~\cite{Supplement}
that if the linear force is weak, $\beta\ll1$, and the inter-band tunneling
is negligible, the center of a wavepacket $x_{c}(t)=\int_{-\infty}^{\infty}\bPsi^{\dag}x\bPsi dx$,
initially chosen as a Bloch wave modulated by an envelope ensuring $\int_{-\infty}^{\infty}\bPsi^{\dag}\bPsi dx=1$,  behaves as  ${\displaystyle dx_{c}/dt=\left(\partial\mu(k)/\partial k\right)_{k=k_0-\beta t}}$. Hence
\begin{eqnarray}
x_{c}(t)=x_{0}+\frac{1}{\beta}[\mu(k_{0} )-\mu(k_{0}-\beta t)]
\label{X_dyn}
\end{eqnarray}
where $x_{0}$ and $k_{0}$ are the initial coordinate and Bloch momentum,
respectively. Since the BZ width is $2$, the wavepacket
crosses the zone after the time interval $2/\beta$.

\begin{figure*}[t]
\includegraphics[width=1\textwidth]{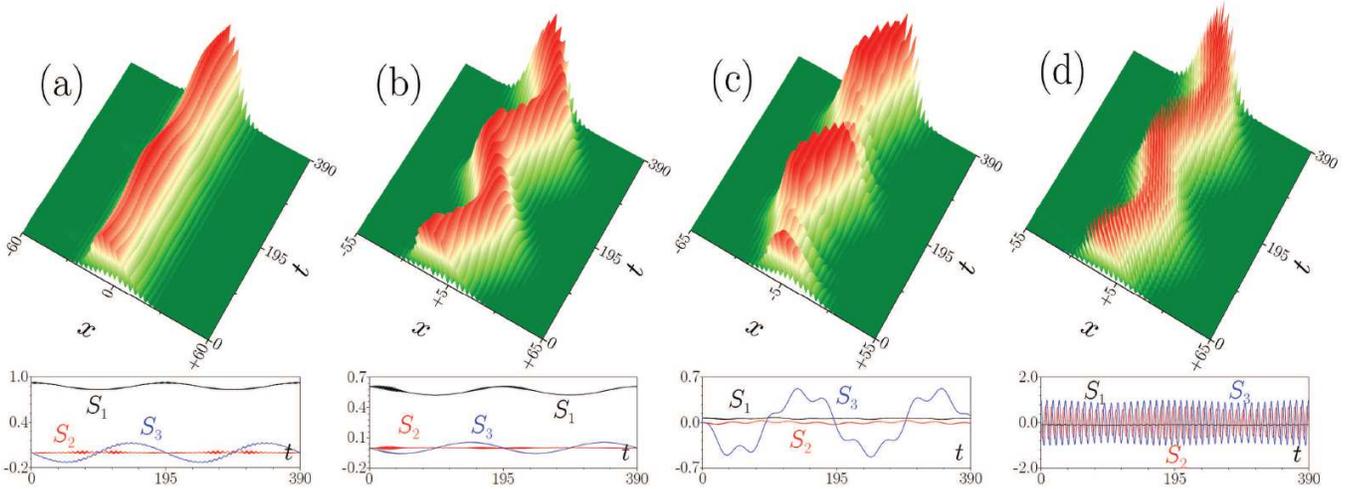}
\caption{(Color online) Dynamics of   $|\Psi_{1}|$
in an OL  for a broad wavepacket, with the width $w_{in}=9\pi/2$), from the second-band
for (a) $\gamma=0.82$, (b) $\gamma=1.85$, (c) $\gamma=4.41$. (d)
Amplitude of the in-phase superposition of the first- and second-band
wavepackets for $\gamma=1.85$ at $\beta=2\Omega_{1}/31\pi$.
The second row shows the evolutions of the spin components.}
\label{fig:three}
\end{figure*}

Important information about the Bloch modes can be obtained
from the symmetries of the Hamiltonian~\cite{com}. Starting with OLs, we
recall~\cite{Lobanov} that the Hamiltonian $H_{{\rm OL}}=H_{0}+V(x)$
obeys the Klein four-group symmetry $\{\hat{1},\alpha_{1},\alpha_{2},\alpha_{3}\}$,
i.e., $[H_{{\rm OL}},\alpha_{j}]=0$, defined by the operators $\alpha_{1}=\sigma_{1}\p$,
$\alpha_{2}=\sigma_{1}\T$, and $\alpha_{3}=\PT$, where $\T$ is the
time reversal operator, $\T\bpsi(x)=\bpsi^{*}(x)$, $\p$ is the parity
operator with respect to $x=0$, i.e., $\p\bpsi(x)=\bpsi(-x)$ and
$\hat{1}$ is the identity operator. Here $\alpha_{i}\alpha_{j}=\alpha_{k}$,
where $i$, $j$ and $k$ are all different. In terms of the Bloch
modes $\boldsymbol{\psi}_{k}$, the operators $\p$ and $\T$ map
a state at $k$ to a state at $-k$: $\p\bu_{k}=\bu_{k}(-x)=\bu_{-k}(x)$
and $\T\bu_{k}=\bu_{k}^{*}(x)=\bu_{-k}(x)$. Thus $\alpha_{1}$ and
$\alpha_{2}$ symmetries imply $k\to-k$ mapping, while $\alpha_{3}$
does not affect $k$. At least one of the eigenfunctions of $H_{{\rm OL}}$
obeys {\em all} $\alpha$-symmetries. When at the boundary
($k=1$) [at the center ($k=0$)] of the BZ
  energy levels do not cross each other, the eigenvalue $\mu(1)$ {[}$\mu(0)${]}
is non-degenerate and its eigenfunction is highly symmetric.  Such
eigenfunction must satisfy $\alpha_{2}\bpsi_{k}=\bpsi_{k}$ with $k=1$
{[}$k=0${]}, and hence $|u_{1,k}|=|u_{2,k}|$. In terms of the average
values of the pseudo-spin (or   spin, for brevity) components
$S_{j}=\int_{-\pi}^{\pi}\bpsi^{\dag}\sigma_{j}\bpsi dx$ ($S_{j}=\int_{-\infty}^{\infty}\bPsi^{\dag}\sigma_{j}\bPsi dx$
for localized nonstationary solutions), this means that at the boundary {[}center{]}
of the BZ $S_{3}=0$. Such phenomenon is observed in the dynamical simulations shown in Fig.~\ref{fig:three}.

In ZLs only one $(\alpha_{3}=\PT)$ of the above
symmetries remains, but there appears an additional one: $\tilde{\alpha}_{1}=\sigma_{1}\tilde{\p}$,
where $\tilde{\p}$ is the reflection with respect to $x=\pi/4$.
Consider an eigenstate at the BZ boundary, $\bpsi^{\mbox{\tiny BZ}}=\bpsi_{k=1}$,
and assume that its eigenvalue is nondegenerate (which implies that
no crossing of the energy levels occurs at $k=1$). It can be chosen to be either
$\alpha_{3}$--symmetric $\bpsi^{\mbox{\tiny BZ}}=\alpha_{3}\bpsi^{\mbox{\tiny BZ}}$,
or satisfying $\tilde{\bpsi}^{\mbox{\tiny BZ}}=\pm\tilde{\alpha}_{1}\tilde{\bpsi}^{\mbox{\tiny BZ}}$.
We set the  ``$+$'' sign (the sign ``$-$'' is analogous).
Non-degeneracy implies the linear dependence $\bpsi^{\mbox{\tiny BZ}}=c\tilde{\bpsi}^{\mbox{\tiny BZ}}$
where $c$ is a complex number, and means that $\bpsi^{\mbox{\tiny BZ}}=\tilde{\alpha}_{1}\bpsi^{\mbox{\tiny BZ}}$
because $\tilde{\alpha}_{1}$ is linear. Thus $\bpsi^{\mbox{\tiny BZ}}$
is simultaneously $\alpha_{3}$ and $\tilde{\alpha}_{1}$-symmetric:
$\bpsi^{\mbox{\tiny BZ}}=\tilde{\alpha}_{1}\bpsi^{\mbox{\tiny BZ}}=\alpha_{3}\bpsi^{\mbox{\tiny BZ}}$.
Using Floquet's theorem we obtain: $\psi_{j}^{\mbox{\tiny BZ}}\left(\frac{\pi}{4}-x\right)=\left[\psi_{j}^{\mbox{\tiny BZ}}\left(x-\frac{\pi}{4}\right)\right]^{*}=\psi_{3-j}^{\mbox{\tiny BZ}}\left(\frac{\pi}{4}+x\right)=-\psi_{j}^{\mbox{\tiny BZ}}\left(-\frac{3\pi}{4}-x\right)$.
The ansatz $x\to x-\frac{\pi}{2}$ from the last equality yields $\psi_{3-j}^{\mbox{\tiny BZ}}\left(x-\frac{\pi}{4}\right)=-\psi_{j}^{\mbox{\tiny BZ}}\left(-\frac{\pi}{4}-x\right)$
for ($j=1,2$), which means that $\bpsi^{\mbox{\tiny BZ}}=-\tilde{\alpha}_{1}\bpsi^{\mbox{\tiny BZ}}$.
This contradicts the assumption that the energy levels
do not cross each other at $k=1$ and that the $\tilde{\alpha}_{1}$-symmetric state
$\bpsi_{k=1}$ is nondegenerate. Therefore, the energy
levels should cross at the boundary of the BZ. Note that  $\tilde{\p}$
reflects the states with respect to the BZ origin (i.e., maps
$k\to-k$) with simultaneous inversion of the spinor components. This
means that $S_{3}$ has opposite signs at the $k$ and $-k$ points.

%[h]

The spectra for OL and ZL are illustrated in Fig.~\ref{fig:one}.
Panels (a) and (b) show transformation of the first $(\mu_{1})$ and
second $(\mu_{2})$ bands of the OL spectrum upon increase of the SO coupling. Panel (c) shows
the two lowest bands $\mu_{1,2}$ of the ZL
at $\gamma=2$. For both types of lattices, increasing SO coupling leads to the appearance
of extrema of $\mu(k)$ in the internal points of the BZ. In OLs, the
slope of the $\mu_{n}(k)$ dependence may change its sign with increasing $\gamma$. According to (\ref{X_dyn}) this means that the SO
coupling may \emph{invert} the direction of motion of the
wavepacket. At a certain $\gamma=\gamma_{f}$ [red lines in Fig.~\ref{fig:one} (a,b)] either the first or the second bands may become
nearly flat. {\color{black} Namely, $\gamma_{f}=1.17, 3.08$ for the first band, and $\gamma_{f}=0.82, 2.93$ for the second band, and note that these values only slightly decrease with growth of the lattice depth}. Such an extreme band-flattening, discussed also in~\cite{YoChu},
implies a weakly dispersive propagation and \emph{almost
complete suppression of BOs\/}. Indeed, according to (\ref{X_dyn})
the amplitude of BOs for an atom in the \emph{n}th band is given by ${\color{black}{\color{black}x_{c}^{{\rm max}}}{\color{black}=[\max_{k}\mu_{n}(k)-\min_{k}\mu_{n}(k)]/\beta}}$
and it must be a  nonmonotonic function of $\gamma$, consistent with
Figs.~\ref{fig:one}(a,b). Band crossing at the edges of the BZ in ZLs is visible in Fig.~\ref{fig:one}(c).

\begin{figure}
%\begin{centering}
\includegraphics[width=1\columnwidth]{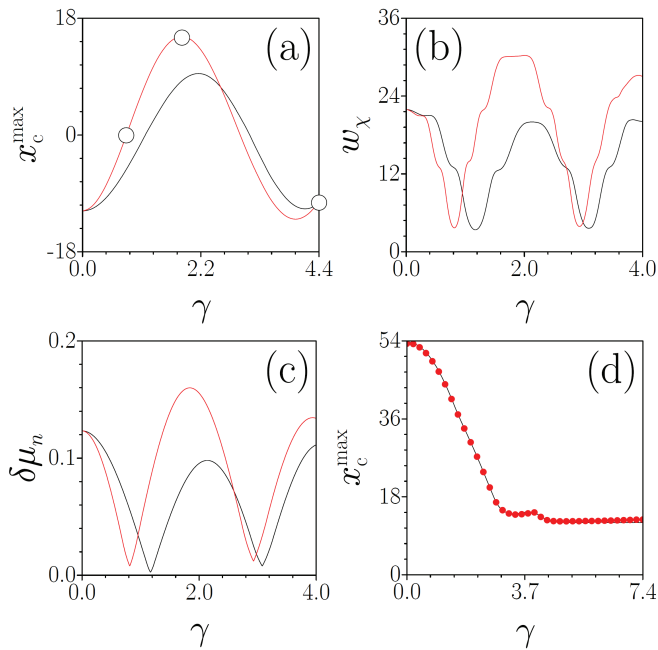}
%\par\end{centering}
\caption{(Color online) (a) Maximal displacement of a broad wavepacket, with a width  $w_{in}=9\pi/2$,
at $k_{0}=0$; (b) Maximal width of a narrow ($w_{in}=\pi/2$)
wavepacket \textit{vs} $\gamma$ in an OL; (c) Width of the  OL   bands
\textit{vs} $\gamma$. The black and red curves correspond to the first
and second bands, respectively. The circles in (a) correspond to the dynamics in Figs.~\ref{fig:three}(a)--(c).
(d) Amplitude of BOs in ZLs  {\em vs} $\gamma$ for a broad wavepacket.
The analytical prediction (red dots) is superimposed to the numerical results (black lines). In all panels $\beta=2\Omega_{1}/31\pi$.}
\label{fig:two}
\end{figure}

The trajectory (\ref{X_dyn}) is in excellent agreement
with the direct solution of the Schr\"odinger equation using a Crank-Nicolson method. Calculations were conducted with $2\times 10^4$  transverse points and steps $dx=0.02$, $dt=0.001$. The input {\color{black} broad wavepacket} had the form  $\bPsi=\bpsi_k(x)\exp(-x^2/w_{in}^2)$, where $\bpsi_k(x)$ is the Bloch wave at $k=0$ for a given $\gamma$.   Fig.~\ref{fig:two}(a)
illustrates the non-monotonic dependence of the maximal displacement $x_{c}^{{\rm max}}$, defined over 20 periods of BOs in an OL. $x_{c}^{{\rm max}}$ becomes zero for specific values
of the SO coupling $\gamma$, where we observe nearly complete
suppression of BOs [Fig.~\ref{fig:three}(a)].
In contrast to broad wavepackets, strongly localized initial states do not experience displacement,
but exhibit {\color{black} nearly} periodic width oscillations in $t$, whose maximal
amplitude $w_{\chi}=\max_{t}\{1/\int_{-\infty}^{\infty}(|\Psi_{1}|^{4}+|\Psi_{2}|^{4})dx\}$
is shown in Fig.~\ref{fig:two}(b) as a function of $\gamma$. Zeros
of $x_{c}^{{\rm max}}(\gamma)$ and minima of $w_{\chi}(\gamma)$
  coincide with the bandwidth mimima from Fig.~\ref{fig:two}(c).

Figures \ref{fig:three}(b,c) show that the
SO coupling may change the direction of the wavepacket motion.
For $\gamma>\gamma_{f}$ and $\gamma<\gamma_{f}$, a wavepacket
initially corresponding to the Bloch mode with $k=0$ moves in the
opposite direction for a given linear force $\beta$.
While the amplitudes of the BOs  in Figs.~\ref{fig:three}(a,b)
differ significantly, the spinor dynamics is similar and is characterized
by the integral spin almost parallel to the SO coupling and exhibiting
weak oscillations in the $(x,z)$-plane: $|S_{1}|\gg|S_{2,3}|$. Increasing the
SO coupling results in rotation of the spin direction, which becomes nearly
orthogonal to the SO coupling. In Fig.~\ref{fig:three}(c) the spin is
mainly directed along the $z$-axis and $|S_{3}|\gg|S_{1,2}|$. Since
flattening of different bands occurs for different values of $\gamma$ {[}Fig.~\ref{fig:two}(c){]},
an arbitrary superposition of several band states always exhibits BOs.

The BOs dynamics changes drastically if the initial wavepacket is a superposition
of states from more than one band. This phenomenon is illustrated in Fig.~\ref{fig:three}(d), which was
obtained for the same parameters as Fig.~\ref{fig:three}(b) but now
for a superposition of eigenmodes from the two lowest bands. Since a two-level
atom is characterized by the Rabi frequency $\Omega_{1}\gg\beta$,
BOs with frequency $\pi\beta$ are accompanied by rapid oscillations between the spinor
components, visible as spin rotation in the plane $(y,z)$, $|S_{1}|\ll1$, whose frequency  exceeds $\Omega_{1}$ because of the effect of the SO coupling.
Such evolution, characterized by two frequencies, can be strictly periodic
only if the frequencies are commensurable, a case that corresponds to a fully
equidistant WSL as discussed below.

\begin{figure}[t]
\begin{centering}
\includegraphics[width=1\columnwidth]{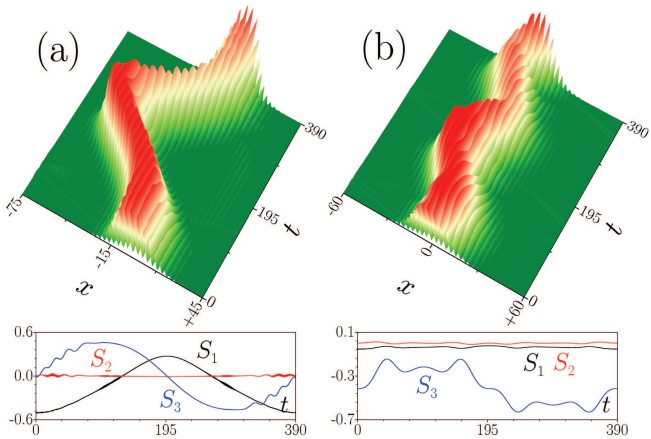}
\par\end{centering}

\caption{(Color online) Dynamics of the amplitude $|\Psi_{1}|$ (upper panels)
and evolution of  the spin components (lower panels) for broad excitations, with
$w_{in}=9\pi/2$, from the first band at (a) $\gamma=1.5$
(b) $\gamma=5.0$ and $\beta=2\Omega_{1}/31\pi$ in a ZL.}
\label{fig:four}
\end{figure}

The BOs dynamics in ZLs  is substantially different than in OLs, due to two main
features of the spectrum: (i) band crossing at BZ edges and (ii) vanishing
of the $S_{3}$ spin-projection at $k=0$ and inversion of $S_{3}$
at $k=\pm1$. Such features have a direct impact on the BOs: the period of
BOs in ZLs  {\em doubles\/} in comparison to that encountered in OLs, and amounts to
$4/\beta$. Indeed, a broad wavepacket starting its motion around $k=-1$
and moving along the lowest band still crosses the BZ after a time period
$2/\beta$, but it arrives to the opposite edge of the BZ at $k=+1$ with
an inverted $S_{3}$ component. Subsequently, it keeps moving along the
\emph{second\/} band, intersecting with the lowest band at $k=1$ and
only after passing the BZ again, but now along a new path, it returns
to the original state after the total time interval of $4/\beta$
(see Fig.~\ref{fig:four}). As in the case of  OLs, for a moderate SO
coupling, the spin dynamics in ZLs is bound to the $(x,z)$ plane ($|S_{1,3}|\gg|S_{2}|)$.
When $\gamma$ grows the spin becomes orthogonal to the SO coupling
{[}$|S_{3}|\gg|S_{1,2}|$ in Fig.~\ref{fig:four}(b){]}. Thus the
amplitude of the BOs in ZLs is determined by the combined width of the \emph{two\/}
lowest bands. This is confirmed by Fig.~\ref{fig:two}(d), which
compares the BOs amplitude obtained numerically with the predictions of (\ref{X_dyn}), but taking the combined width of the two lowest
bands as a total range of variation of $\mu$. The BOs amplitude
in ZLs  becomes independent of $\gamma$ for a sufficiently strong SO
coupling, because one of the spinor components can be strongly
suppressed at $\gamma\gg1$ and the dynamics resembles that of the one-component
system, {\color{black}as explained in~\cite{LiYe}}).

\begin{figure}%[h]
\begin{centering}
\includegraphics[width=1\columnwidth]{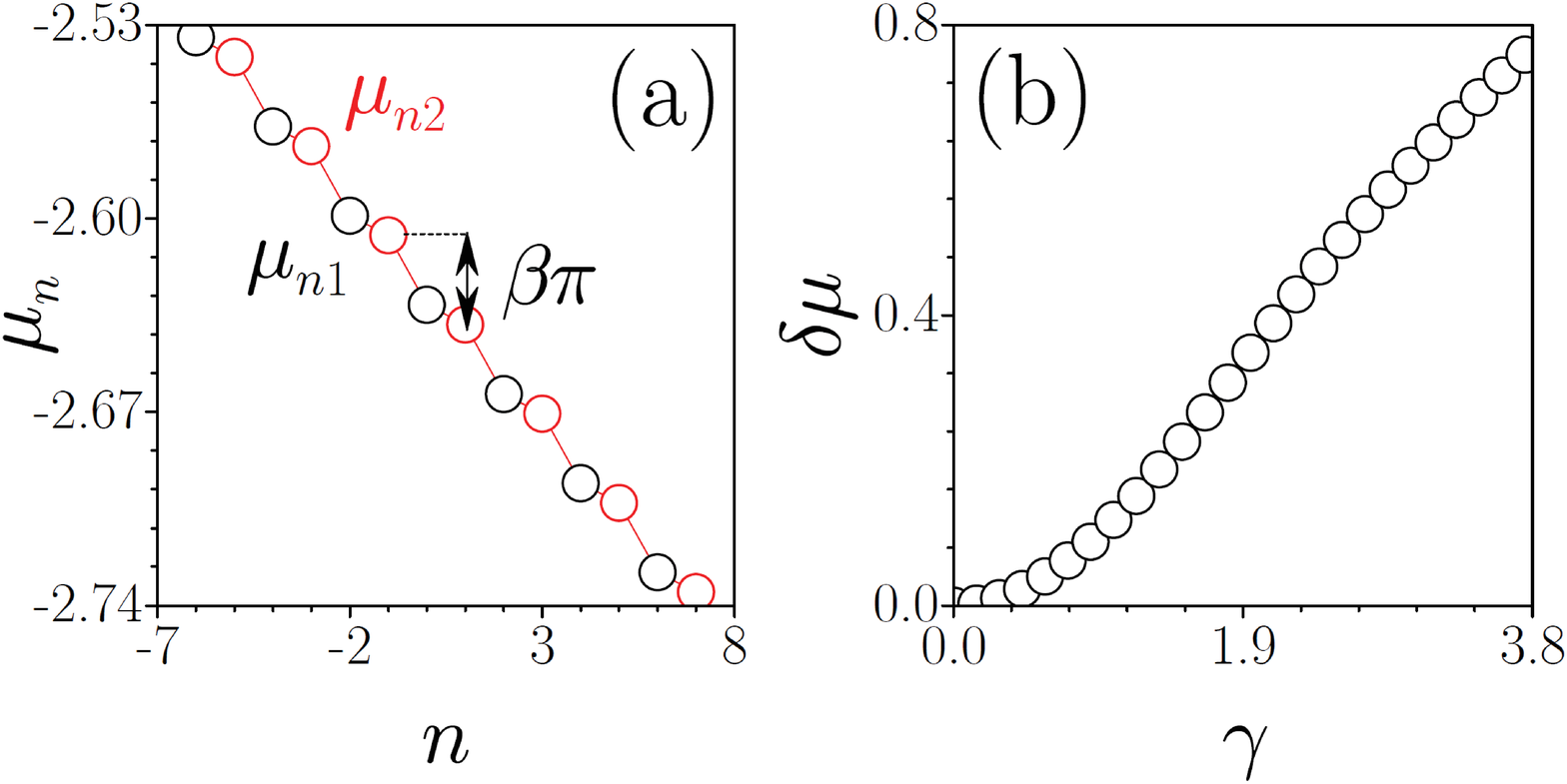}
\par\end{centering}

\caption{(Color online) (a) Part of the WSL of an OL   with $\gamma=2$ and
$\beta=2\Omega_{1}/31\pi$. (b) Mutual shift of two sub-ladders in
the spectrum, as a function of $\gamma$. }
\label{fig:five}
\end{figure}

An alternative approach to analyze the BOs is based on the calculation of the \emph{localized\/} Wannier-Stark modes of the total
Hamiltonian $H_{0}+H_{L}+\beta x$ on numerically large windows with zero boundary
conditions. For OLs  such a spectrum ($\mu_{n}$ {\it vs} eigenvalue
number $n$) is shown in Fig.~\ref{fig:five}(a). It reveals a
  specific WSL with \emph{two\/} characteristic steps $\mu_{n+1}-\mu_{n}$
between neighboring eigenvalues. Due to the spinor character of the system
the total WSL is a combination of two equidistant sub-ladders, namely $\mu_{n_{1}}$
and $\mu_{n_{2}}$, shifted along $\mu$. At $\gamma=0$ the mutual
shift of the sub-ladders amounts to $2\Omega_{1}$. The Wannier-Stark modes from each
sub-ladder have identical first and second components, and  they
are in-phase for one sub-ladder and out-of-phase in the other sub-ladder.
If BOs and Rabi periods are commensurable, i.e., $2\Omega_{1}=m\beta\pi$
at $\gamma=0$, the two sub-ladders overlap exactly, the separation of the neighboring
eigenvalues becomes equal to $\beta\pi$, and \emph{any} excitation
is recovered after the period $2/\beta$. At $2\Omega_{1}=(m+1/2)\beta\pi$
the \emph{entire\/} ladder becomes equidistant with $\mu_{n+1}-\mu_{n}=\beta\pi/2$,
and the period of BOs for any wavepacket exciting modes from \emph{both\/} sub-ladders
increases to $4/\beta$. At $\gamma\neq0$ the SO coupling results in an
additional mutual shift $\delta\mu$ of the two sub-ladders shown in Fig.~\ref{fig:five}(b).
Due to such a shift the dynamics is in general non-periodic for
inputs that excite \emph{both} sub-ladders
unless the SO coupling is selected such that $\delta\mu=m\pi\beta/2$
($m\in\mathbb{N}$) and the equidistance of the entire spectrum is restored. If only
\emph{one\/} sub-ladder is excited the dynamics may still be periodic.
In contrast to the behavior observed in  OLs, the WSL for  ZLs  is equidistant
for any $\gamma$ with $\mu_{n+1}-\mu_{n}=\beta\pi/2$, which  indicates BOs with a period $4/\beta$.

Summarizing, we have shown that SO coupling brings important new features into
Bloch oscillations. It allows controlling the direction of atomic motion
and enables nearly complete suppression of oscillations in OLs. In ZLs, oscillations feature period-doubling due to crossing of energy bands at the boundary of the Brillouin zone.

%\smallskip{}

\begin{acknowledgments}
The work was supported by the FCT (Portugal) grant
UID/FIS/00618/2013, by the Severo Ochoa Excellence program (Spain), and by Fundacio Cellex Barcelona.
\end{acknowledgments}

\section*{Supplemental Material for \textit{``Bloch oscillations in optical and Zeeman lattices in the presence
		of spin-orbit coupling''}}

Following the ideas of~\cite{Zener,Houston}, in this Supplemental Material
we outline the derivation   of the  relation
\begin{eqnarray}
\frac{dx_{c}}{dt}=\left.\frac{\partial\mu_{\alpha_{0}}(k)}{\partial k}\right|_{k=k_0-\beta t},\label{xc_dyn}
\end{eqnarray}
which %for $B(t)=\beta t$ 
leads to Eq.~(1) from the main text.  In fact, we will address   a slightly more general case: we consider the situation when the linear force $\beta$ depends on time: $\beta=\beta(t)$. The particular case $\beta(t)=\textrm{const}$ obviously corresponds to the situation addressed in the main text. The derivation
is valid for state described by  a ket-vector $|\bPsi\rangle$ of any dimension (i.e. having
one, two or more components) and for
the arbitrary strength of the SO-coupling $\gamma$ (provided the Conditions formulated below are not violated).

We use the complete set of Bloch states $|k,\alpha\rangle$
of the Hamiltonian $H$: $H|k,\alpha\rangle=\mu_{\alpha}(k)|k,\alpha\rangle$,
where $\mu_{\alpha}(k)$ is the energy of the state, the Bloch quasi-momentum
$k$ belongs to the reduced Brillouin zone (BZ): $k\in(-1,1]$, and
$\alpha=1,2,...$ is the band number ($\alpha=1$ being the number
of the lowest band). Following Houston's approach~\cite{Houston},
we introduce adiabatically-varying Bloch states $|\kappa (t),\alpha\rangle$
where
\begin{eqnarray}
\kappa (t)=k-B(t),\qquad B(t)=\int_{0}^{t}\beta(t')dt',\label{B(t)}
\end{eqnarray}
and $k$ is the initial value of the quasi-momentum (or central
quasi-momentum in the case of localized Bloch wavepackets). The spinor
$|\Psi\rangle$ can be expanded in terms of the adiabatically-varying states as:
\begin{eqnarray}
|\Psi\rangle=\sum_{\alpha=1}^{\infty}\int_{-1}^{1}dk\chi_{\alpha}(k,t)|\kappa(t),\alpha\rangle.\label{expan_new}
\end{eqnarray}
and, for the sake of convenience, the spectral coefficients will be
represented as
\begin{eqnarray}
\chi_{\alpha}(k,t)=\chi_{\alpha}^{(0)}(k,t)e^{-i\int_{0}^{t}\mu_{\alpha}[\kappa(\tau)]d\tau}\label{chi_1}
\end{eqnarray}
with the functions $\chi_{\alpha}^{(0)}(k,t)$ to be determined latter.
The normalization condition (in the direct and Fourier spaces)
\begin{eqnarray}
\langle\Psi|\Psi\rangle=\sum_{\alpha=1}^{\infty}\int_{-1}^{1}|\chi_{\alpha}(k,t)|^{2}dk=1\label{normal}
\end{eqnarray}
is also imposed.

We assume that the following conditions hold:

\paragraph{Condition 1:}

Bloch states of only one band, say of the band $\alpha_{0}$, are
initially excited, i.e. $\chi_{\alpha}(k,0)=0$ for $\alpha\neq\alpha_{0}$.

\paragraph{Condition 2:}

A wavepacket $|\Psi\rangle$ is a Bloch wave with a smooth and broad envelope
(compared to the lattice period) and its spectrum centered
at a quasi-momentum $k_{0}$ in the reduced BZ is much
narrower than the BZ zone, so that the approximation
\begin{eqnarray}
\int_{-1}^{1}dk|\chi_{\alpha_0}(k,t)|^{2}\frac{\partial\mu_{\alpha_0}(k)}{\partial k}\approx\frac{\partial\mu_{\alpha_0}(k_{0})}{\partial k_{0}}%\int_{-1}^{1}|\chi_{\alpha}(k,t)|^{2}dk
%=\frac{\partial\mu_{\alpha}(k_{0})}{\partialk_{0}}
\label{eq:approx_mu}
\end{eqnarray}
is valid [here we take into account the normalization (\ref{normal})].

\paragraph{Condition 3:}

The linear force is weak, i.e., $|\beta(t)|\ll1$, so that inter-band
tunneling is negligible.

Let us start with the case $\beta\equiv0$.
%
%in the Schr\"odinger equation $i\mathbf{\boldsymbol{\Psi}_{\mathit{t}}}=(H_{0}+H_{L})\mathbf{\boldsymbol{\Psi}}$,  and
Using the expansion
\begin{eqnarray}
|\Psi\rangle=\sum_{\alpha=1}^{\infty}\int_{-1}^{1}dk\chi_{\alpha}(k,t)|k,\alpha\rangle\label{expan}
\end{eqnarray}
where
\begin{eqnarray}
\chi_{\alpha}(k,t)=\chi_{\alpha}^{(0)}(k)e^{-i\mu_{\alpha}(k)t},
\label{chi_0}
\end{eqnarray}
from the Schr\"odinger equation  $i\mathbf{\boldsymbol{\Psi}_{\mathit{t}}}=(H_{0}+H_{L})\mathbf{\boldsymbol{\Psi}}$
one gets
\begin{eqnarray}
\label{X_dot}
i\frac{dx_c}{dt} & = &
\langle\Psi|[x,H_0+H_L]|\Psi\rangle\nonumber \\
& = & \sum_{\alpha,\alpha'=1}^{\infty}\int_{-1}^{1}\!dk'\int_{-1}^{1}\!dk\chi_{\alpha'}^{*}(k',t)\chi_{\alpha}(k,t)\nonumber \\
&  & \times\,(\mu_{\alpha}(k)-\mu_{\alpha'}(k'))\langle k',\alpha'|x|k,\alpha\rangle.
\end{eqnarray}
Using   Floquet theorem, the Bloch states can be expressed in the
form of the  expansion
\begin{eqnarray}
|k,\alpha\rangle=e^{ikx}\sum_{n=-\infty}^{\infty}\bc_{\alpha,n}(k)e^{2inx}
\label{Fourier}
\end{eqnarray}
where $\bc_{\alpha,n}(k)$ are vector Fourier coefficients. The Bloch
states are  orthonormal, which   implies
\begin{eqnarray}
\sum_{m=-\infty}^{\infty}\bc_{\alpha',m}^{\dag}(k)\bc_{\alpha,m}(k)=\delta_{\alpha\alpha'}.
\label{Perceval}
\end{eqnarray}
Now, using (\ref{Fourier}) we get % \begin{widetext}
\begin{eqnarray}
\langle k',\alpha'|x|k,\alpha\rangle%\int_{-\infty}^{\infty}\!\!dx\!\!\sum_{m,n=-\infty}^{\infty}\bc_{\alpha',m}^{\dag}(k')\bc_{\alpha,n}(k)xe^{i(k-k'+2n-2m)x}
%=
%
%
%
=-2\pi i\!\!\sum_{m,n=-\infty}^{\infty}\bc_{\alpha',m}^{\dag}(k')\bc_{\alpha,n}(k)\nonumber \\
\times\frac{\partial}{\partial k}\delta(k-k'+2n-2m),
\end{eqnarray}
%\end{widetext}%Since $k,k'\in(-1,1]$ we have that $-2<k-k'<2$ and all terms in (\ref{delta}) with $m\neq n$ are zero. Thus and
and using the relation
$
-i\langle k',\alpha'|\frac{\partial}{\partial k}|k,\alpha\rangle%\quad\quad\ensuremath{}
$
we obtain
\begin{eqnarray}
\label{aux1}
\langle k',\alpha'|\frac{\partial}{\partial k}H|k,\alpha\rangle%
=\langle k',\alpha|\frac{\partial}{\partial k}\mu_{\alpha}(k)|k,\alpha\rangle\nonumber \\
=\frac{\partial\mu_{\alpha}(k)}{\partial k}\delta_{\alpha\alpha'}\delta(k-k')+\mu_{\alpha}(k)\langle k',\alpha|\frac{\partial}{\partial k}|k,\alpha\rangle.
\end{eqnarray}
On the other hand, the same expression is computed as
\begin{eqnarray}
\langle k',\alpha'|\frac{\partial}{\partial k}H|k,\alpha\rangle%=\langlek',\alpha'|H\frac{\partial}{\partialk}|k,\alpha\rangle%
=\mu_{\alpha'}(k')\langle k',\alpha'|\frac{\partial}{\partial k}|k,\alpha\rangle\label{aux2}
\end{eqnarray}
From (\ref{aux1}) and (\ref{aux2}) we obtain the useful relation
\begin{eqnarray}
\label{useful}
(\mu_{\alpha'}(k')-\mu_{\alpha}(k))\langle k',\alpha|\frac{\partial}{\partial k}|k,\alpha\rangle\nonumber \\
=\frac{\partial\mu_{\alpha}(k)}{\partial k}\delta_{\alpha\alpha'}\delta(k-k').
\end{eqnarray}
Combining (\ref{useful}) %with (\ref{auxil1})
and (\ref{X_dot})
we obtain
\begin{eqnarray}
\frac{dx_{c}}{dt}=
\sum_{\alpha=1}^{\infty}\int_{-1}^{1}dk|\chi_{\alpha}(k,t)|^{2}\frac{\partial\mu_{\alpha}(k)}{\partial k}.
\label{X_dot_1}
\end{eqnarray}
Equation (\ref{X_dot_1}) is exact, no approximations were made so
far. Using now the Conditions 1 and 2  formulated above,
one arrives at the formula
\begin{eqnarray}
\frac{dx_{c}}{dt}=\frac{\partial\mu_{\alpha_{0}}(k_{0})}{\partial k_{0}}.
\label{velocity}
\end{eqnarray}

Let us now turn to the case when the field gradient is present, $\beta(t)\neq0$. Following Houston's approach~\cite{Houston},
we introduce adiabatically varying Bloch states $|\kappa (t),\alpha\rangle$, i.e. consider (\ref{Fourier}) with substitution $k\to\kappa(t)$,
where
\begin{eqnarray}
\kappa (t)=k-B(t),\qquad B(t)=\int_{0}^{t}\beta(t')dt',
\label{B(t)}
\end{eqnarray}
By direct differentiating and using that at each instant $t$, $|\kappa(t),\alpha\rangle$
are the eigenfunctions of $H$ with $\mu_{\alpha}(\kappa(t))$ we
obtain
\begin{eqnarray}
i\frac{\partial|\Psi\rangle}{\partial t}=H|\Psi\rangle+\beta(t)x|\Psi\rangle+i|F\rangle\label{eq:3}
\end{eqnarray}
where
\begin{eqnarray*}
	|F\rangle=\sum_{\alpha=1}^{\infty}\int_{-1}^{1}\!dk\left[\frac{\partial\chi_{\alpha}^{(0)}(k,t)}{\partial t}\bpsi_{\alpha,k}-\beta(t)\chi_{\alpha}^{(0)}(k,t)\frac{\partial\bpsi_{\alpha,k}}{\partial\kappa}\right]\\
	\times\exp\left[-i\int_{0}^{t}\mu_{\alpha}(\kappa(\tau))d\tau\right]e^{i\kappa(t)x}dk
\end{eqnarray*}

Now we require $\chi_{\alpha}^{(0)}(k,t)$, which, so far, are arbitrary
functions of time, to ensure $|F\rangle=0$. To this end we require $\langle\kappa,\alpha|F\rangle=0$
for all $\alpha$, which yields the set of equations
\begin{eqnarray*}
	\frac{\partial\chi_{\alpha}^{(0)}(k,t)}{\partial t}=\beta\sum_{\alpha'=1}^{\infty}\int_{-1}^{1}dk'\chi_{\alpha'}^{(0)}(k',t)e^{i\Delta_{\alpha'\alpha}(\kappa',\kappa)}\\
	 \times\int_{-\infty}^{\infty}dxe^{i(k'-k)x}\bpsi_{\alpha,\kappa}^{\dag}(x)\frac{\partial}{\partial\kappa'}\bpsi_{\alpha',\kappa'}(x),
\end{eqnarray*}
where $\kappa'=k'-B(t)$ and
\begin{eqnarray}
\Delta_{\alpha'\alpha}(\kappa',\kappa)=\int_{0}^{t}\left[\mu_{\alpha}'(\kappa'(\tau))-\mu_{\alpha}(\kappa(\tau))\right]d\tau\label{Delta}
\end{eqnarray}
As we already did above, now we use the Fourier expansion (\ref{Fourier})
with the substitution $k\to\kappa(t)$ and obtain
\begin{eqnarray}
\frac{\partial\chi_{\alpha}^{(0)}(k,t)}{\partial t}=\beta(t)\sum_{\alpha'=1}^{\infty}\chi_{\alpha}^{(0)}(k,t)e^{i\Delta_{\alpha'\alpha}(\kappa,\kappa)}\nonumber\\
\times\sum_{n=1}^{\infty}\bc_{\alpha',n}^{\dag}(\kappa)\frac{\partial}{\partial\kappa}\bc_{\alpha,n}(\kappa). \label{eq:4}
\end{eqnarray}
This is an exact formal result representing an infinite number of
equations. If however all three conditions introduce above  hold,
Eq.~(\ref{eq:4}) for a given band $\alpha$ can be approximated by
\begin{equation}
\frac{\partial\chi_{\alpha}^{(0)}(k,t)}{\partial t}\approx\beta(t)\sum_{\alpha'=1}^{\infty}\chi_{\alpha'}^{(0)}(k,t)\frac{\partial}{\partial\kappa}\sum_{n=1}^{\infty}\bc_{\alpha',n}^{\dag}(\kappa)\bc_{\alpha,n}(\kappa).\label{eq:5}
\end{equation}
Since at $\beta=0$ we have the conservation (\ref{Perceval}), we
conclude that
\begin{eqnarray}
\frac{\partial}{\partial\kappa}\sum_{n=1}^{\infty}\bc_{\alpha,n}^{\dag}(\kappa)\bc_{\alpha,n}(\kappa)=\mathcal{O}(\beta)
\end{eqnarray}
and thus
%\begin{eqnarray}${\partial\chi_{\alpha}^{(0)}(k,t)}/{\partial t}=\mathcal{O}(\beta^{2})$
%\end{eqnarray}i.e. is negligible in the leading approximation. Thus,
one can use (\ref{velocity}) with substitution $k_{0}\to k_{0}-B(t)$,
leading to Eq.~(\ref{xc_dyn}) at $\beta=$const.

\end{document}